\documentclass[]{aastex63}

\usepackage{url} 
\usepackage{graphicx}
\usepackage{float}
\usepackage{amssymb,amsmath}
\graphicspath{{Figs/}}
\DeclareGraphicsExtensions{.png,.pdf}

\received{\today}
\revised{xx. xx, xxxx}
\accepted{xx. xx, xxxx}
\submitjournal{ApJ}

\shorttitle{Identifying magnetic reconnection in 2D-HVM simulations with CNN}
\shortauthors{Hu et al.}
\graphicspath{{./}{Figs/}}

\begin{document}

\title{Identifying magnetic reconnection in 2D Hybrid Vlasov Maxwell simulations with Convolutional Neural Networks}

\correspondingauthor{Andong Hu}
\email{andong@cwi.nl}

\author[0000-0002-6929-2158]{A. Hu}
\affiliation{Centrum Wiskunde \& Informatica, Amsterdam, The Netherlands, EU}

\author{M. Sisti}
\affiliation{Dipartimento di Fisica, Universit\`a di Pisa, Italy, EU}
\affiliation{Aix-Marseille University, CNRS, PIIM UMR 7345, Marseille, France, EU}

\author{F. Finelli}
\affiliation{Dipartimento di Fisica, Universit\`a di Pisa, Italy, EU}

\author{F. Califano}
\affiliation{Dipartimento di Fisica, Universit\`a di Pisa, Italy, EU}

\author{J. Dargent}
\affiliation{Dipartimento di Fisica, Universit\`a di Pisa, Italy, EU}

\author{M. Faganello}
\affiliation{Aix-Marseille University, CNRS, PIIM UMR 7345, Marseille, France, EU}

\author{E. Camporeale}
\affiliation{Centrum Wiskunde \& Informatica, Amsterdam, The Netherlands, EU}
\affiliation{CIRES, University of Colorado, Boulder, CO, USA}
\affiliation{NOAA Space Weather Prediction Center, Boulder, CO, USA}

\author{J. Teunissen}
\affiliation{Centrum Wiskunde \& Informatica, Amsterdam, The Netherlands, EU}



\begin{abstract}

Magnetic reconnection is a fundamental process that quickly releases magnetic energy stored in a plasma. Identifying, from simulation outputs, where reconnection is taking place is non-trivial and, in general, has to be performed by human experts. Hence, it would be valuable if such an identification process could be automated. Here, we demonstrate that a machine learning algorithm can help to identify reconnection in 2D simulations of collisionless plasma turbulence. Using a Hybrid Vlasov Maxwell (HVM) model, a data set containing over 2000 potential reconnection events was generated and subsequently labeled by human experts. We test and compare two machine learning approaches with different configurations on this data set. The best results are obtained with a convolutional neural network (CNN) combined with an `image cropping' step that zooms in on potential reconnection sites. With this method, more than 70\% of reconnection events can be identified correctly. The importance of different physical variables is evaluated by studying how they affect the accuracy of predictions. Finally, we also discuss various possible causes for wrong predictions from the proposed model. 

\end{abstract}



\section{Introduction}
\label{sec:introduction}

Magnetic reconnection is a fundamental process in space and laboratory plasmas in which magnetic energy is converted into kinetic energy, released in the form of accelerated particles, flows and heating \citep{cassak2007scaling}. Although the process itself is highly localized, it eventually leads to a global change of the magnetic field topology.

Reconnection typically occurs in the presence of thin, elongated current sheets (CSs), which can locally become unstable but eventually re-arrange the connectivity of magnetic field lines on a global scale. This process is forbidden at large fluid scales (with respect to kinetic and/or diffusive scales) where ideal MHD holds. At such scales, the initial connectivity of field lines is preserved since field lines are ``frozen-in'' in the fluid motion of the plasma (and vice versa). Local violation of ideal MHD laws leads to the onset of reconnection \citep{Furth1963, COPPI1979370, White_1980}.


In space, magnetic reconnection is as of today recognized as the energetic driver of several important energetic processes as, for instance, solar flares and coronal mass ejections~\citep{Priest_1982}.
It also occurs routinely at the dayside boundary between the solar wind and the Earth's magnetosphere as well as in the magnetotail. As a consequence, accelerated particles are injected into the magnetosphere in some cases down to the Earth polar regions \citep{PriestForbes_2000, dungey1961interplanetary}. Magnetic reconnection is therefore behind many of the risks associated with space weather, including electronic damage to satellites, endangering astronauts, disturbing Global Navigation Satellite System signals and even impacting power grids \citep{cassak2016}.
Magnetic reconnection also occurs in conditions where CSs are naturally created by the presence of global large-scale unsteady flows, as for dayside or tail reconnection. Furthermore, CSs can be created by the development of MHD-scale vortices driven by the Kelvin-Helmholtz instability along the magnetospheric flanks \citep{faganello2017} or by the non linear dynamics of magnetic field fluctuations such as small scale vortex motion in solar wind turbulence \citep{retino2007,phan2018, haynes2014reconnection}. 

Magnetic reconnection plays a key role in the context of plasma turbulence, a phenomenon today routinely observed by satellites in the solar wind and in the Earth's  magnetosphere. With respect to a turbulent fluid where energy is transferred by wave-wave interactions, reconnection has been recognized to represent an alternative path for energy transfer in plasmas \citep{Cerri2017,karimabadi2013coherent,camporeale2018coherent}. First, the local formation of current sheets (CSs) efficiently transfers energy from the large magnetohydrodynamic (MHD) scales to the small ion kinetic scale (which governs the thickness of CSs). Second, CS disruption allows to inject energy directly on the sub-ion scales involving also the electrons strictly coupled to the magnetic small scale dynamics. These dynamics are generic and visible in almost all simulations of plasma turbulence.


In 2D, the CS structure is characterized by the presence of a thin in-plane magnetic field inversion region associated with an out-of-plane directed current. It corresponds to a magnetic field that points in the opposite direction after crossing the so-called neutral line where the in-plane field goes to zero. In the presence of a nearly constant out-of-plane magnetic field dubbed ``guide field'', the plasma is in the so-called ``guide field regime''. In this regime the non linear dynamics is dominated by in-plane interactions (quasi-2D regime). This is the regime adopted in the present paper. We underline that in 2D reconnection can more easily be identified by human experts than in 3D. Another advantage is that simulations are computationally much cheaper in 2D. The use of 2D simulations therefore  allows us, as a first step, to carefully assess the viability of the proposed machine learning approach.

\begin{figure}
  \centering
  \includegraphics[width=0.32\textwidth]{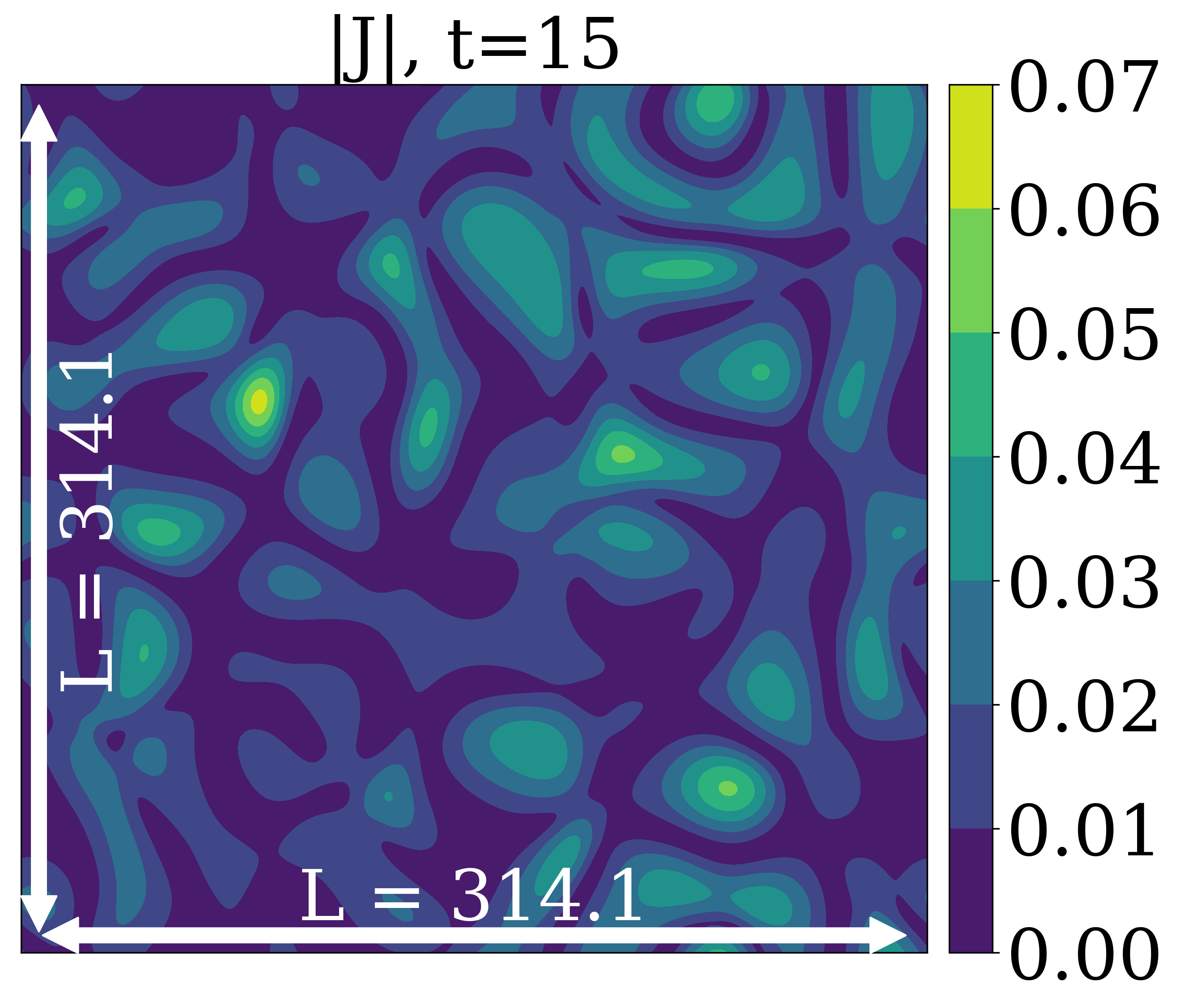}
  \includegraphics[width=0.32\textwidth]{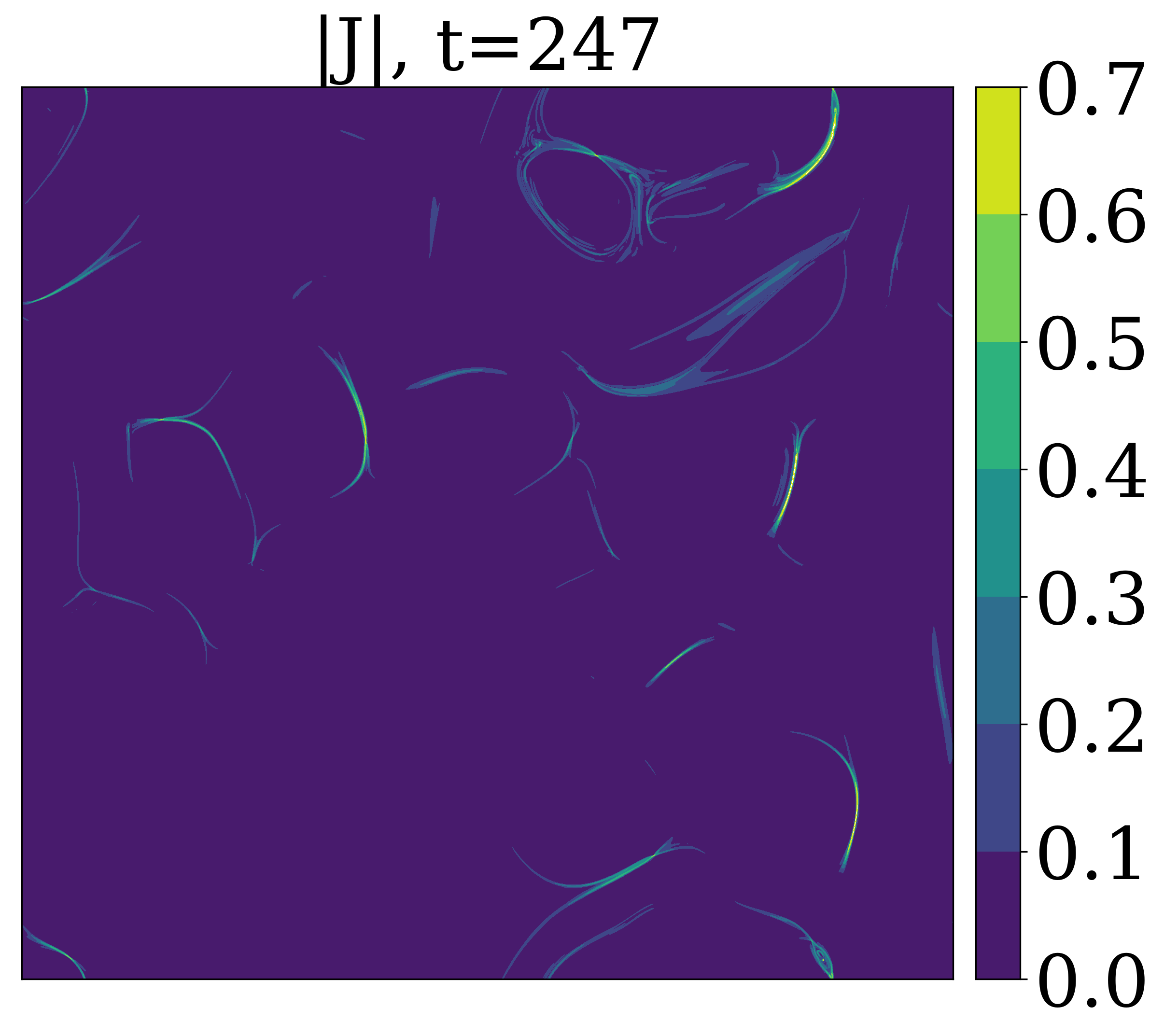}
  \includegraphics[width=0.32\textwidth]{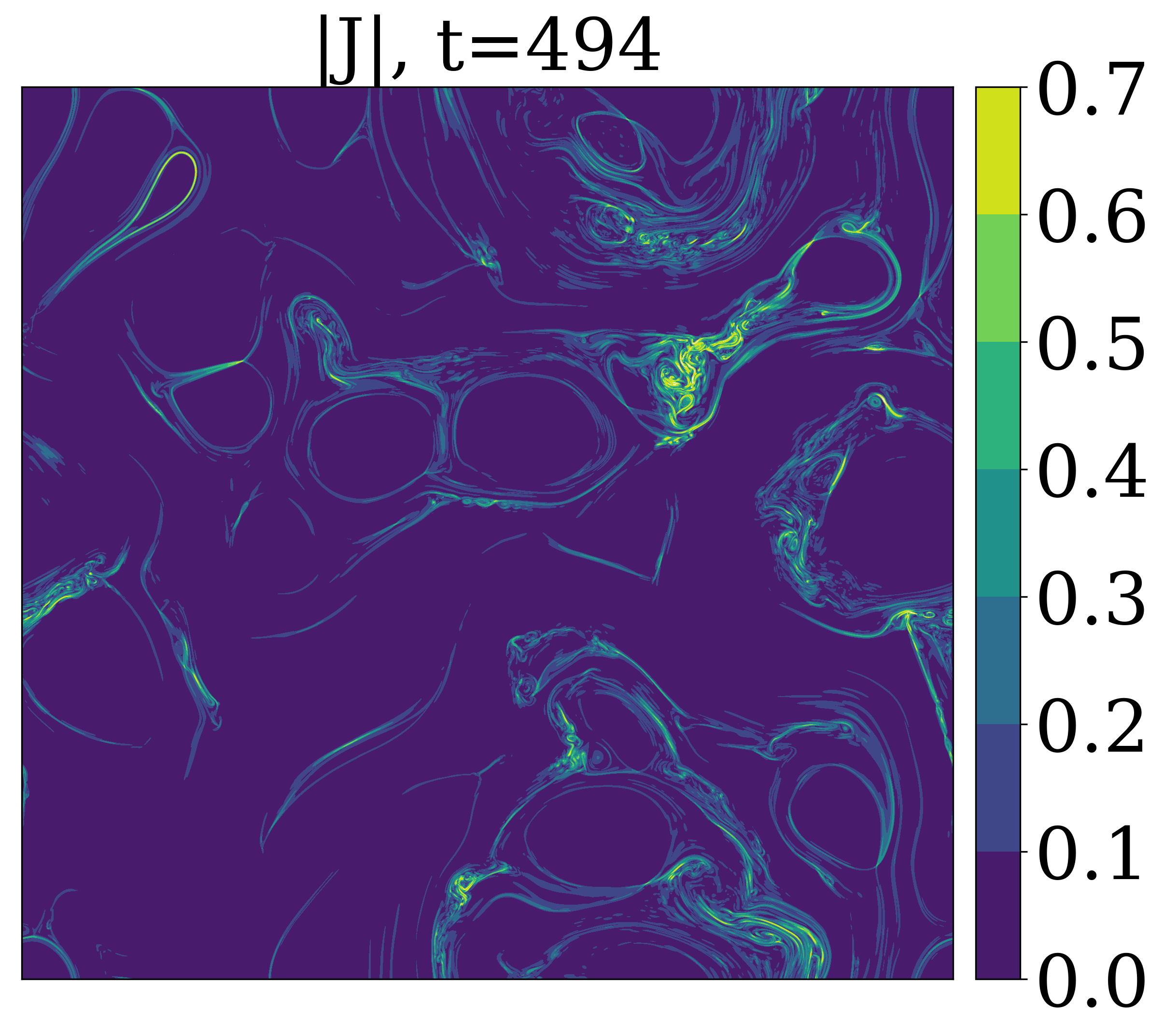}
  \caption{Evolution of the current density $|\mathbf{J}|$ in Sim 1, see Table \ref{tab:data sets}. Time is indicated in units of the inverse ion cyclotron frequency $\Omega_{ci}^{-1}$. At $t=15$, the beginning of the simulation, the initial perturbations are visible. At $t=247$, which corresponds to roughly one eddy turnover time ($t_e\sim 260$), current sheets have formed. They show up as thin and elongated peaks in the current density. At $t=494$, turbulence is fully developed and current sheets are broken up by the dynamics at small scales.}
  
  \label{fig:current_time}
\end{figure}

An example of a 2D kinetic Vlasov simulation of a turbulent magnetized plasma is shown in Fig. \ref{fig:current_time}. In this simulation (presented in detail in Section \ref{sec:simulation}), turbulence is generated by initial large-amplitude magnetic perturbations of wavelengths of the order of the domain size. After approximately one eddy turnover time for the largest-wavelength perturbations (at $t=247\,\Omega_{ci}^{-1}$, where $\Omega_{ci}$ is the the ion cyclotron frequency, see section \ref{sec:simulation}), we see that after an initial transient thin CSs (with respect to the domain size) form. After another eddy turnover time, at $t = 494\,\Omega_{ci}^{-1}$ the plasma is in a fully turbulent regime where the CSs interact, merge or disrupt. It becomes harder to distinguish individual CSs, but reconnection continues to occur.

An increasing amount of data for the study of magnetic reconnection is continuously produced by simulations and by satellite measurements. It is therefore important to find ways to reliably and efficiently locate reconnection events in these data. Recognizing magnetic reconnection is relatively straightforward in idealized configurations that are prepared ad-hoc, such as a symmetric isolated CS usually modeled in  simulations as a 1D equilibrium (the so-called Harris sheet; see, e.g., \cite{camporeale2005model}). However, in more realistic dynamical configurations, detecting reconnection is much less trivial as for instance in the context of large-scale vortex dynamics \citep{Daughton2014,borgogno2015,sisti2019satellite}
or in turbulent simulations \citep{Servidio2009,ZhdankinAPJ2013}.
As of today, there is no single optimal way to automatically identify magnetic reconnection in simulations of ``non idealized''
plasma configurations.  With observational data this task is even more difficult, since most signatures of reconnection, such as large-amplitude field variations or particle accelerations, can be caused by other phenomena (e.g., large-scale vortices, plasma turbulence, shocks).
Up to now, reconnection events have  been identified by human experts, which is time-consuming and can lead to subjective results. Therefore, a method that can automatically and reliably identify reconnection events would be valuable. 

Machine learning techniques have recently been used to identify regions with reconnection based on signatures in the particle velocity distribution function~\citep{Dupuis_2020}. In this work we instead focus on the electromagnetic, density and velocity fields to detect magnetic structures where reconnection is occurring using supervised machine learning techniques.
More specifically, we have developed a method capable of identifying reconnection in two-dimensional simulations of plasma turbulence performed with a Hybrid Vlasov Maxwell (2D-HVM) numerical code~\citep{valentini2007, valentini2014hybrid}. The simulations are described in section \ref{sec:simulation}.
Compared to a more demanding fully kinetic description, a proper description of the electron reconnection physics is lacking here. Nevertheless the use of a hybrid model has several advantages. A hybrid model is able to simulate a physical domain much larger than the ion kinetic scale, which is a typical scale for the thickness of current sheets. A simulation can therefore contain lots of current sheets, while still resolving ion kinetic effects for a proper descriptions of turbulence at the ion/current sheet scale.
There are also several reasons for using 2D simulations.
As mentioned above, reconnection can clearly be identified in 2D, whereas in 3D reconnection is not clearly defined and much harder to identify, and computational costs are much lower in 2D. Also,  although 2D plasma turbulence is geometrically simpler than its 3D counterpart, it still creates a large number of magnetic field geometries where reconnection can occur in ``non-idealized'' configurations. Furthermore, magnetic reconnection can be identified more reliably by human experts in 2D data than in 1D (time series) observational data. Simulations provide more information as compared to a satellite's 1D time series since they contain a full 2D description of the plasma dynamics and fields shaping around a reconnection zone. Using simulations can be seen as a first step in the automatic identification of reconnection. Finally, additional data can readily be generated in the future for refining the method. The long-term objective is to apply machine learning methods to time series data, such as those provided by a virtual satellite technique, and real satellite measurements.

In this paper we use supervised machine learning methods, which means that a training data set consisting of input-output pairs is required. After the training phase, the model can predict outputs for other unseen inputs. The inputs here consist of simulation data in a $200^2$ pixels neighborhood around a potential reconnection zone, and the output is a binary value indicating whether reconnection is taking place. It should be noted that a $200^2$ pixel area is able to capture most important information around the reconnection site and it is convenient for human experts for labeling the samples. A challenge in this binary classification problem is that the input is relatively high-dimensional. Several machine learning approaches are considered. In the simplest one, we extract only a few statistics from the input data, and then apply a so-called decision tree classifier. The other classifiers that we use are based on neural networks.

Neural networks (NNs) are widely used for machine learning. Non-linear NNs contain few assumptions on the `basis functions' used to map inputs to outputs, in contrast to e.g.~a linear model. In principle, NNs of sufficient complexity can approximate any continuous function mapping inputs to outputs.
Their flexibility makes NNs a powerful tool for space physics modeling, and they have been applied in various contexts, e.g.~forecasting geomagnetic indices \citep{gruet2018multiple, camporeale2018machine,wu1996prediction,wu1997geomagnetic, wing2005kp, wintoft2017forecasting,bala2009real} and modeling for the upper atmosphere \citep{hu2018using,hu2019Modeling,hu2020deep, hoque2011new}.
The use of NNs is attractive here because the precise relationship between magnetic reconnection and selected physical parameters is not entirely known.

Here, we use convolutional neural networks (CNNs) which are particularly well suited for dealing with images. A CNN employs a type of filtering that can extract spatial features at a given characteristic scale, while retaining spatial transformation invariance. The repeated application of these filters can process the input image on a number of different scales and at different levels of feature abstraction. Many of the recent breakthroughs in image recognition have been achieved with CNNs. In a space context, CNNs have been used, for example, for space object detection \citep{linares2016space} and solar flare prediction \citep{ park2019generation, huang2018deep}.

The paper is organized as follows. In Section \ref{sec:simulation}, the 2D-HVM model and the simulation data are introduced. The generation of the reconnection data set is described in section \ref{sec:samples-labeling}, and the proposed machine learning models and an image cropping method are introduced in section \ref{sec:ml-methods}. The accuracy of the machine learning models is assessed in Section 3, by comparing the performance of the different methods on unseen data. The contribution of individual physical variables is analyzed in Section \ref{sec:discussion-conclusion}, where we also study an optimal window size for the image cropping method. The importance of each variable is also investigated in this section. Several illustrative examples of misclassifications are also discussed in  Section \ref{sec:discussion-conclusion}. Finally, our results are summarized in Section \ref{sec:summary-outlook} where we also discuss their relevance for the automatic classification of real observational data as e.g.~those from the Magnetospheric Multiscale Mission (MMS).

\section{Methods and Data}
\label{sec:method}

\subsection{Simulations}
\label{sec:simulation}
Data are provided by means of high-resolution 2D Hybrid Vlasov-Maxwell (HVM) simulations of turbulence. In this model ions are fully kinetic and electrons are modeled as a neutralizing fluid with mass through a generalized Ohm’s law \citep{valentini2007,perrone2012vlasov}. Quasi neutrality, $n_i \simeq n_e \simeq n$, is assumed. 
Then, the  system of equations is given by the Vlasov equation for the ion distribution function $f_i=f_i({\bf x},{\bf v},t)$
\begin{equation}\label{eq:vlasov}
\frac{\partial f_i}{\partial t}+{\bf v}\cdot\nabla f_i + ({\bf E}+ {\bf v}\times{\bf B})\cdot\frac{\partial f_i}{\partial {\bf v}}=0
\end{equation}
where ${\bf E}$ and {$\bf B$} are the electric and magnetic field. The generalized Ohm's equation for the electron response reads
\[
{\bf E}-d_e^2\nabla^2{\bf E}=-({\bf u}\times{\bf B})+\frac{1}{n}({\bf J}\times{\bf B}) 
-\frac{1}{n}\nabla P_e 
\]
\begin{equation}\label{eq:ohm}
+ \frac{d_e^2}{n}\nabla\cdot[{\bf uJ}+{\bf Ju}]-\frac{1}{n}d_e^2\nabla\cdot(\frac{{\bf JJ}}{n})
\end{equation}
Furthermore, the Faraday and Ampere equations are given by
\begin{equation} \label{eq:farampere}
\frac{\partial {\bf B}}{\partial t}=-\nabla\times{\bf E} \,; \;\;\;
\nabla\times{\bf B}={\bf J}
\end{equation}
where the displacement current has been neglected (low-frequency regime). The ion density $n$ and the ion fluid velocity ${\bf u}$ are obtained by taking the zeroth and first order velocity moment of $f_i$, respectively. All equations are normalized to the ion mass $m_i$, the initial ion cyclotron frequency $\Omega_{ci}= e B_0/m_i c$, where $B_0$ is the magnitude of the initial guide field along the $z$-direction, and the Alfvén velocity $v_A$ (or, equivalently, to the ion skin depth $d_i = v_A\Omega_{ci}^{-1}$). As a result, the electron skin depth is given by $d_e = \sqrt{m_e/m_i}$, where $m_e$ is the electron mass.
We assume an isothermal equation of state for the electron pressure, $P_e = nT_{0e}$. The set of equations (\ref{eq:vlasov}-\ref{eq:farampere}) is solved
in a 2D-3V phase space using an Eulerian algorithm \citep{MangeneyJCP2002} which combines the so-called splitting scheme with the current advanced method \citep{valentini2007}.

We use data from two simulations, which are listed in Table \ref{tab:data sets}.
In both simulations we take $B_z(t=0) = B_0=1$ and $B_x(t=0) = B_y(t=0) = 0$, where the $z$-direction is perpendicular to the simulation plane.
Random isotropic magnetic-field perturbations are added to the initial equilibrium configuration to initiate turbulence.
The plasma response self-consistently generates velocity and density fluctuations. The initial perturbations have wavenumber magnitudes $k\in[0.02,0.12]$ and a root mean squared value of the magnetic fluctuations equal to $\delta B_{rms} \simeq 0.3$ in both simulations.

The initial distribution functions are Maxwellian with a uniform initial temperature $T_{0i} = T_{0e}$ such that the ion beta parameter 
$\beta_{\rm i}\doteq 2 n T_{i0}/B^2_0$ is equal to one. The velocity space is sampled by $51^3$ uniformly distributed grid points spanning $[-5v_{th,i},5v_{th,i}]$ in each direction, where $v_{th,i}=\sqrt{\beta_{i}/2}$ is the initial ion thermal velocity. We set the reduced mass ratio to $m_{\rm i}/m_{\rm e}=100$ so that $d_{i}$ and $d_{e}$ are separated by one decade.


\begin{table}
  \caption{Description of the two 2D-HVM simulations used in this paper. Both simulations are performed on a square $L\times L$ domain with $L=50 \times 2\pi$. The resolution of Sim 1 is higher, as indicated by its smaller grid spacing $dl$. $N_\mathrm{samples}$ is the number of labeled samples; in the training, validation and test data sets only non-ambiguous samples are included. The computational costs of Sim 1 and Sim 2 were about $5$ M and $2$  M core hours, respectively.}
  \begin{tabular}{c|c|ccccccc}
   Name & description & grid size & $dl / d_i$ & $N_\mathrm{samples}$ & \% reconnection & time range ($\Omega_{ci}^{-1}$)\\
    \hline
    Sim 1 & all data & $3072^2$ & $0.1$ & 2069 & 42 \% & [0, 370] \\
    & training set & & & 1205 & 34.7 \% & [0, 260], [340, 370] & & \\
    & validation set & & & 437 & 56 \% & [280, 320] & & \\
    \hline
    Sim 2 & test set & $2048^2$ & $0.15$ & 124 & 56.5 \% & [205, 233]
  \end{tabular}
  \label{tab:data sets}
\end{table}

\subsection{Extraction and labeling of images from simulations}
\label{sec:samples-labeling}

\begin{figure}
  \centering
  \includegraphics[width=0.5\textwidth]{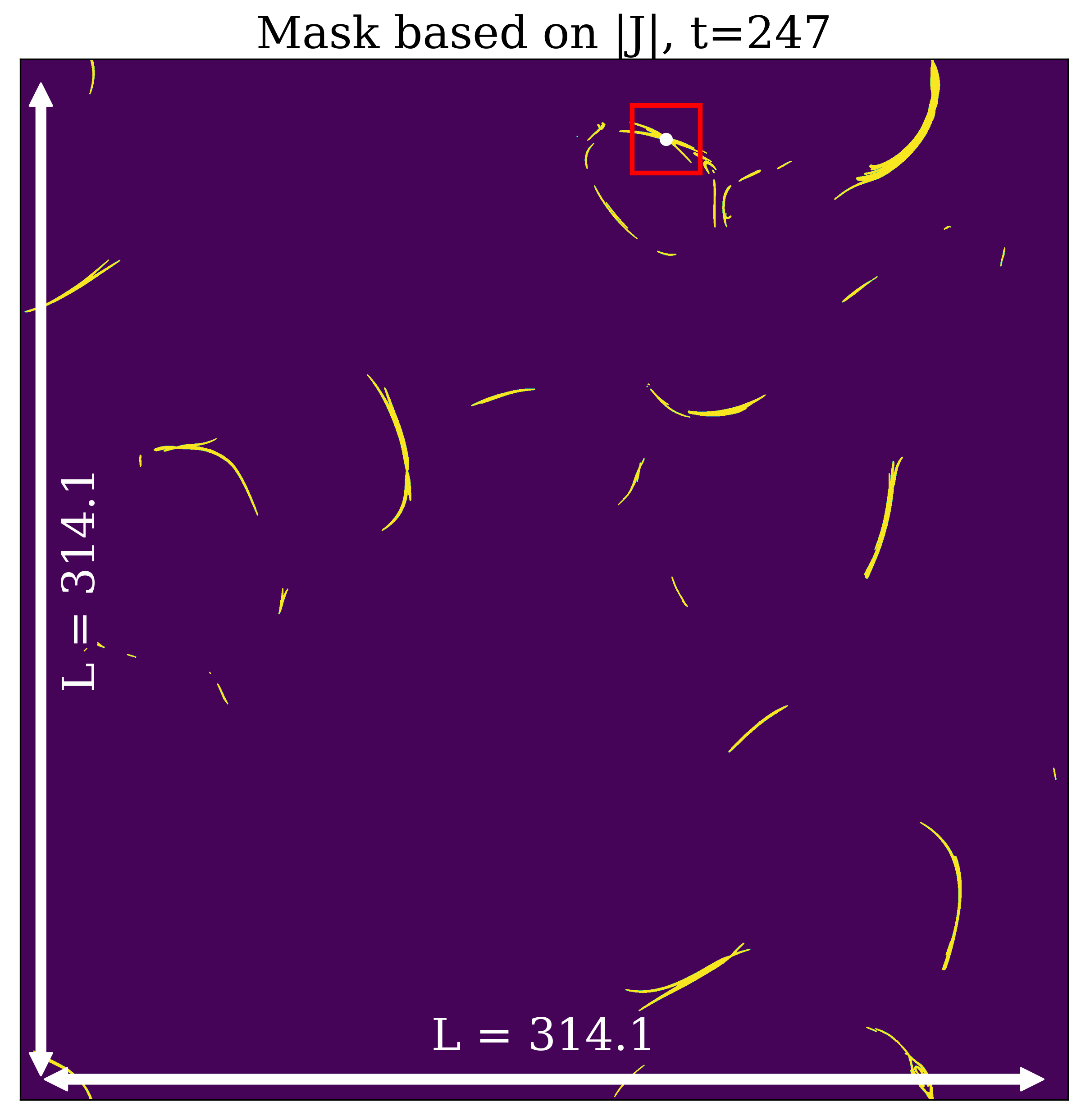}
  \caption{An example showing how potential reconnection sites in the simulation output are selected. First, a mask is constructed based on a threshold for the current density (shown in yellow). A list of current sheets is extracted, which are centered on their local peaks (e.g., the red square centered on the white dot). Note that there are typically multiple reconnection events that occur simultaneously}
  \label{fig:Jn}
\end{figure}

\begin{figure}
  \centering
  \includegraphics[width=1.0\textwidth]{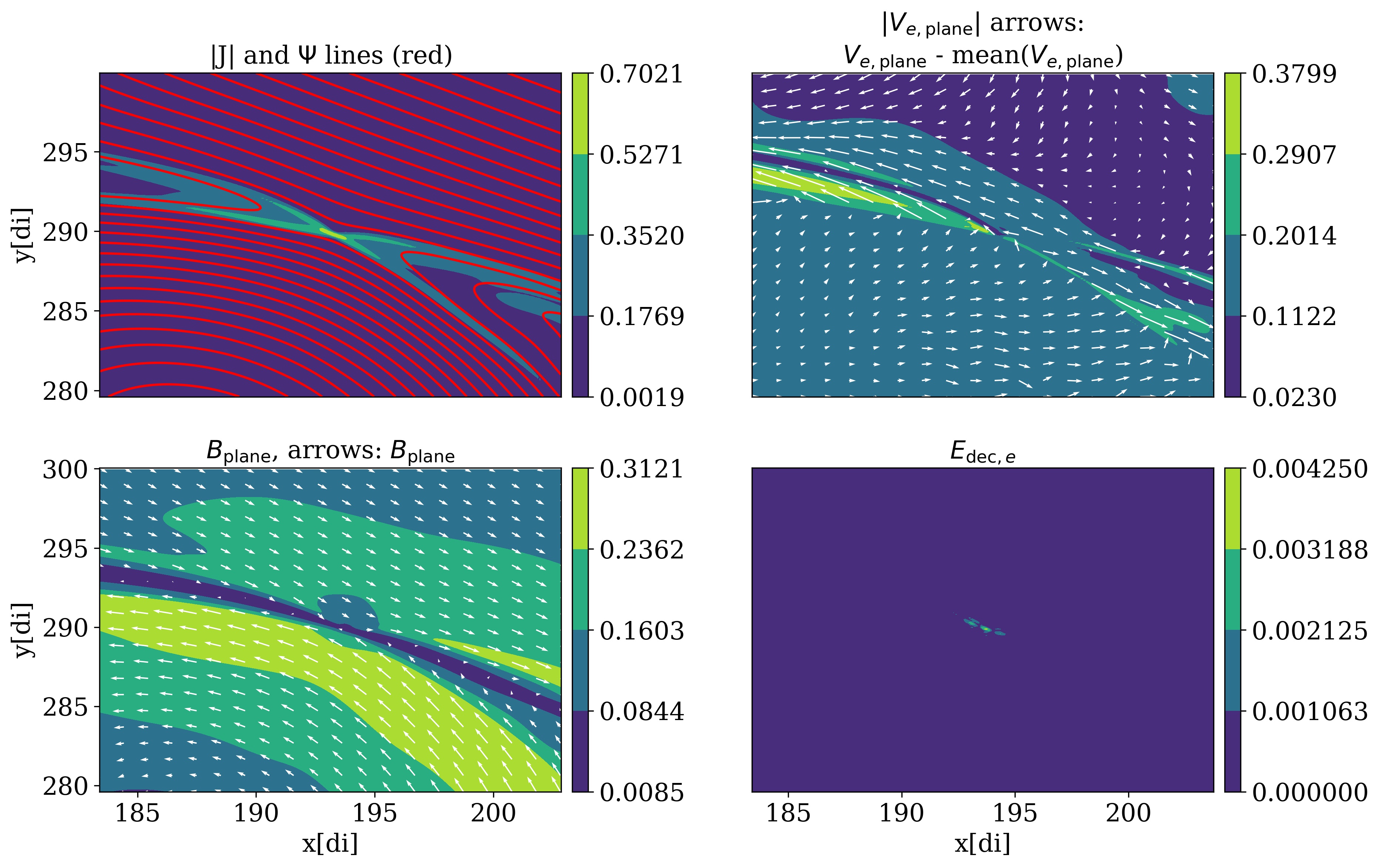}
  \caption{An example of the pictures used for the classification of magnetic reconnection by human experts. The selected variables $|{\bf J}|$, $\Psi$, $V_{e, \mathrm{plane}}$, $B_\mathrm{plane}$ and $E_{\mathrm{dec}, e}$ that can be indicators for reconnection are shown. Note that arrows corresponding to ${\bf V}_{e,\mathrm{plane}}$ and ${\bf B}_\mathrm{plane}$ help human classification but only the magnitudes of these vectors is used in the machine learning models. This example corresponds to the red box drawn in Fig. \ref{fig:Jn}. }
  \label{fig:Mask}
\end{figure}

\begin{figure}
  \centering
  \includegraphics[width=0.8\textwidth]{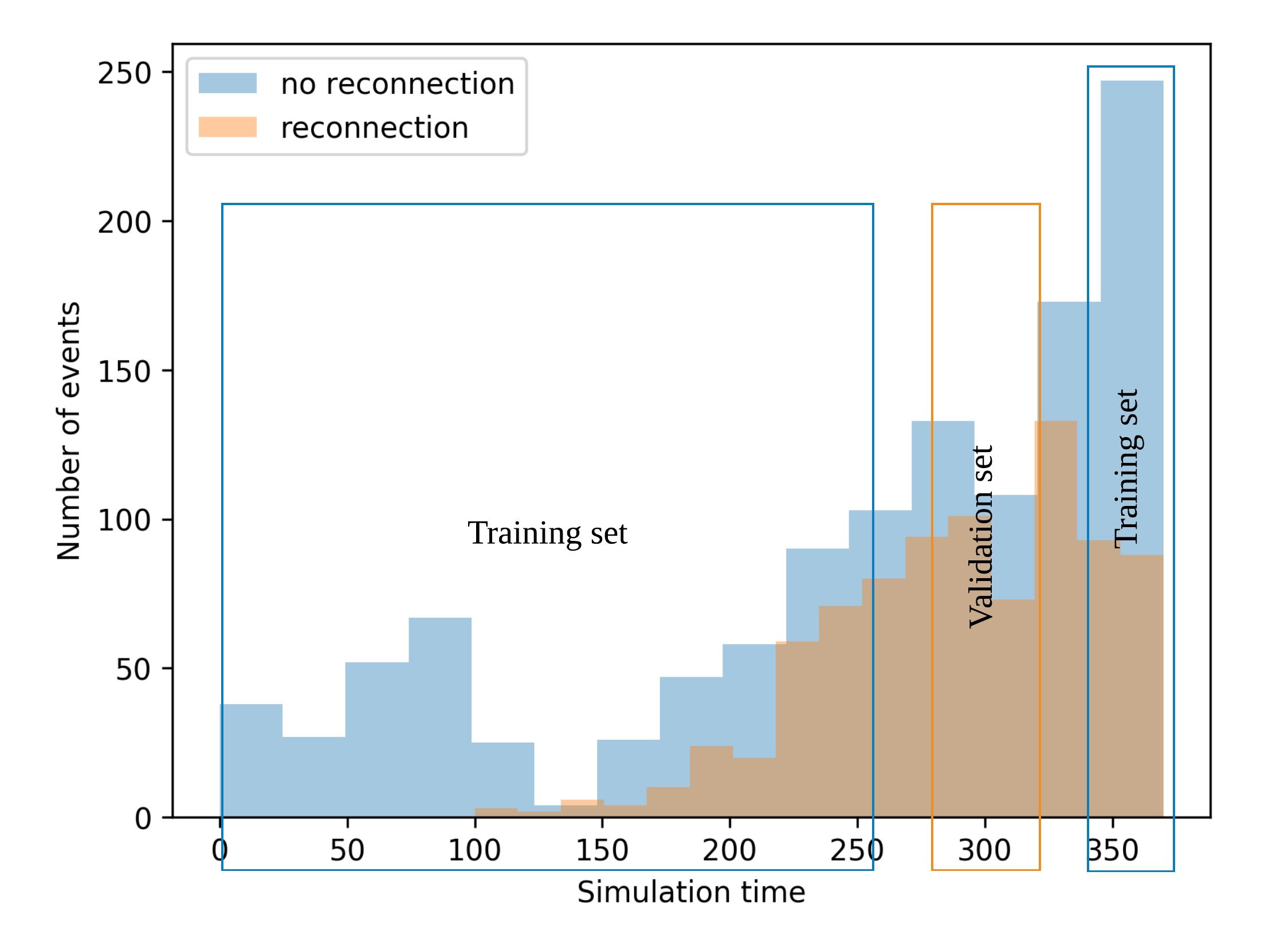}
  \caption{Distribution of the samples and their labels of Sim 1. The blue rectangles indicate the time range for the training set, and the yellow rectangle the time range for the validation set. A gap of $\delta t = 20$ is used between the training and validation set to keep them independent.}
  \label{fig:true-false-vs-time}
\end{figure}

Magnetic reconnection usually takes place in or nearby a maximum of the current density. The current density can therefore be used to extract potential reconnection sites from the simulation data. For each time at which simulation output was saved, all regions are marked where the magnitude $J = |{\bf J}|$ of the current density exceeds a given threshold. The threshold is defined by $J_{th}=\max(\sqrt{<J^2>+3\sigma(J^2)})$, where $\sigma(J^2)=\sqrt{<J^4>-(<J^2>)^2}$ \citep{Zhdankin2013} where the brackets denote the average over the whole physical domain. 
The center of each reconnection site is determined by each local maximum of the current density. Around a local maximum one finds a long and narrow region where the current density exceeds the threshold. 
For such a region, we first determine the maximal distance $d_\mathrm{max}$ to the local maximum. Then a square of size $(2 \,  d_\mathrm{max})^2$ is extracted, centered on the local maximum. 

An example of a potential reconnection site is highlighted by a red square box in Fig. \ref{fig:Jn}. In some cases multiple potential reconnection sites can be present in the same region. All these regions are extracted and together they constitute the samples of the data set. All samples are rescaled to $200^2$ pixels to facilitate their use in machine learning methods. An example of the extracted data is shown in Fig. \ref{fig:Mask}.  The data extracted from the simulation outputs used in the machine learning models contain seven physical variables that are listed below.
\begin{itemize}
  \item Current magnitude $J=|{\bf J}|$
  \item Electron fluid velocity along z direction ${V}_{e, z} = {u}_z - {J}_z/ne$
  \item In-plane electron fluid velocity $V_\mathrm{plane} = \sqrt{{V}_{e,x}^2 + {V}_{e,y}^2}$, where ${V}_{e, x/y} = {u}_{x/y} - {J}_{x/y}/(ne)$
  \item In-plane magnetic field $B_\mathrm{plane} = \sqrt{B_x^2 + B_y^2}$
  \item Magnetic field fluctuation along the z-direction $\delta B_z = B_z - \langle B_z \rangle$
  \item Flux function $\Psi$, in 2D, related to the magnetic field through ${\bf B}_\mathrm{plane} = \nabla \Psi \times {\bf z}$
  \item Electron decoupling defined by 
  $E_{\mathrm{dec}, e} = |({\bf E} + {\bf V}_e \times {\bf B})_z|$, which corresponds to the $z$-component of the electric field in the electron rest frame. In 2D, it is non-zero only when the magnetic field dynamics and the electron motion are decoupled, which is a necessary condition for reconnection to occur.
\end{itemize}

We remark that the ion decoupling term $E_{\mathrm{dec}, i}= |({\bf E} + {\bf u} \times {\bf B})_z|$ is included in the pictures for classification by a human expert, but it is not used as a variable for the machine learning models. There are several reasons for this. First, our experts use $E_{dec,i}$ only as an auxiliary variable to $E_{dec,e}$. Since $E_{dec,i}$ is more sensitive to large scale structures than $E_{dec,e}$, it is heavily influenced by the environment around the candidate site, making it less useful. Furthermore, the ML method will only benefit from additional variables if enough training samples are available. We `only' have around 3000 labeled samples, and leaving out some rather uninformative variables can be beneficial.



As listed in Table \ref{tab:data sets}, two data sets have been generated. Sim 1 is the main one including 2069 samples. Sim 1 is divided into two subsets: a training set and a validation set. The training set is used to train the model and the validation set is used to find the optimal parameters for the trained model. A time interval of 20 $\Omega_{ci}^{-1}$ is used between these two data sets to keep them independent, as shown in Fig. \ref{fig:true-false-vs-time}. This delay is important to prevent information from leaking from the training to the validation set, since it takes some time for the morphology of reconnection sites to change. Sim 2 is used as the test set to assess the accuracy of the developed models as well as their performance on simulation data with a different resolution than the one used for training. 

Once a sufficient number of $200^2$ images have been extracted from the simulations they need to be labeled by human experts. We make use of an automated workflow on \url{zooniverse.org}, which is a platform aimed at involving the general public in the labeling of scientific data sets. The project can be accessed via \url{http://aida-space.eu/reconnection}, together with a tutorial on how to identify reconnection sites. The project is public, so any expert can help with the labeling.

\begin{figure}
  \centering
  \includegraphics[width=0.6\textwidth]{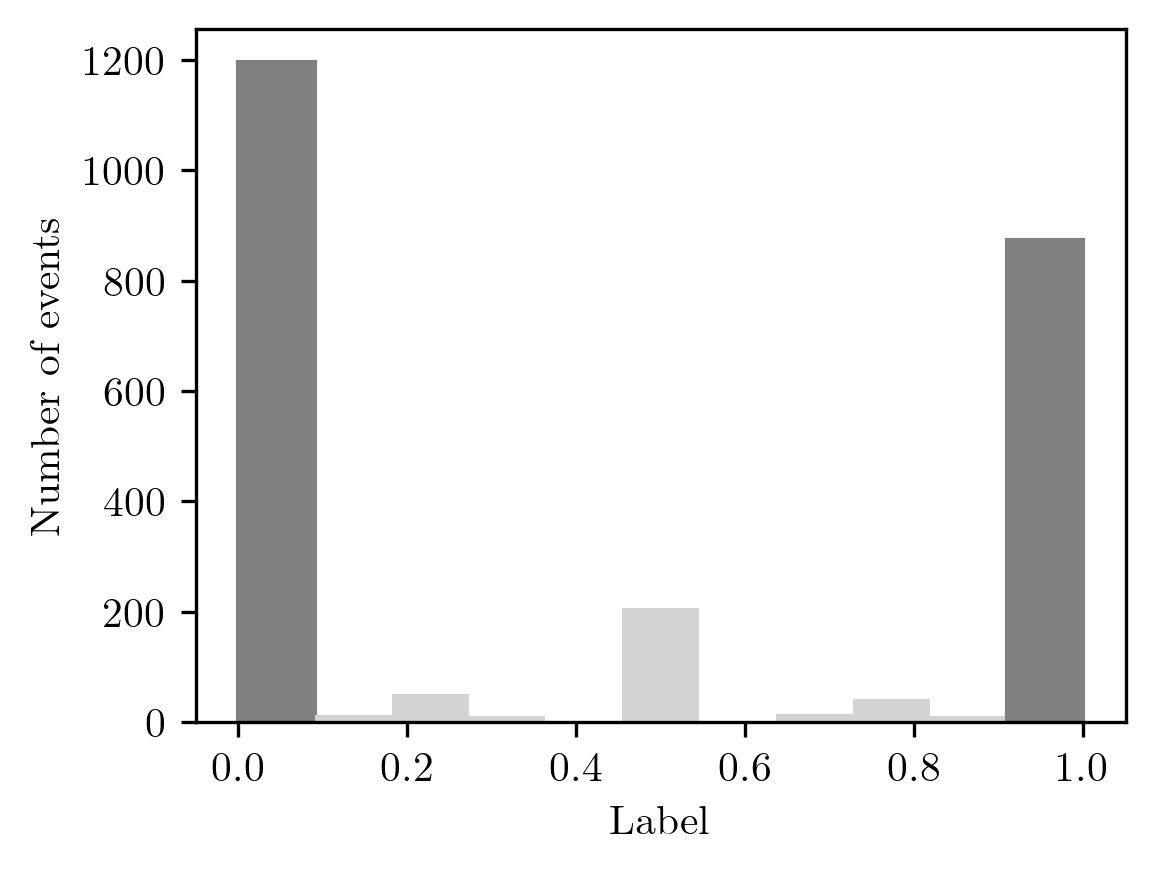}
  \caption{Distribution of labels in Sim 1, see table \ref{tab:data sets}.  Here $1$ indicates reconnection, $0$ indicates no reconnection, and values in between are ambiguous cases. Note that the data set is fairly balanced. In this study, only non-ambiguous events (indicated by the darker shade) are used to train and evaluate the machine learning models.}
  \label{fig:label-statistics}
\end{figure}

It is worth noticing that magnetic reconnection in ``non-idealized'' configurations as those created by turbulence can be difficult to identify, even for experts. Therefore, the possible labels assigned to a picture are: 1 for reconnection, 0 for no reconnection, and 0.5 for ambiguous cases.
Every picture is labeled by up to three human experts, after which the labels are averaged, so that in the end each case has a single label. The distribution of labels in Sim 1 is shown in Fig. \ref{fig:label-statistics}. Labels unequal to zero or one are considered to be ``ambiguous''. There are around 1200 negative cases, 900 positive cases and 300 ambiguous cases. To convert these labels to binary values all ambiguous cases with a label unequal to 0 or 1 were dropped. However, this does not mean all included cases are completely unambiguous. At the time of writing,  about 59\% of cases was labeled by one human expert, 32\% by two experts and 9\% by three experts. The corresponding fractions of ambiguous cases were 8\%, 25\% and 30\%, showing that the number of ambiguous cases goes up significantly when cases are inspected by multiple experts. From these numbers, we can estimate that roughly 14\% of cases in our filtered data set would still be marked ambiguous, if all cases had been labeled by three experts. For the smaller Sim 2 data set all cases were labeled by three experts, so the above estimate does not apply. In future research, it would be interesting to include ambiguous events in the data set. One could then investigate whether such events can be identified by a machine learning approach, and it would also allow for a direct comparison between the performance of a machine learning model and that of a human expert.

\subsection{Machine Learning Approaches}
\label{sec:ml-methods}

We consider two machine learning models. \textbf{CNN-X} takes a region of size $X^2$ as input. This subset is determined by a physics-based image cropping approach. The whole image is used when $X=200$. A \textbf{decision tree} classifier is used for comparison. These models are described in more detail below. 

\subsubsection{CNN-X}
\label{sec:feature-selection}

\begin{figure}
  \centering
  \includegraphics[width=0.9\textwidth]{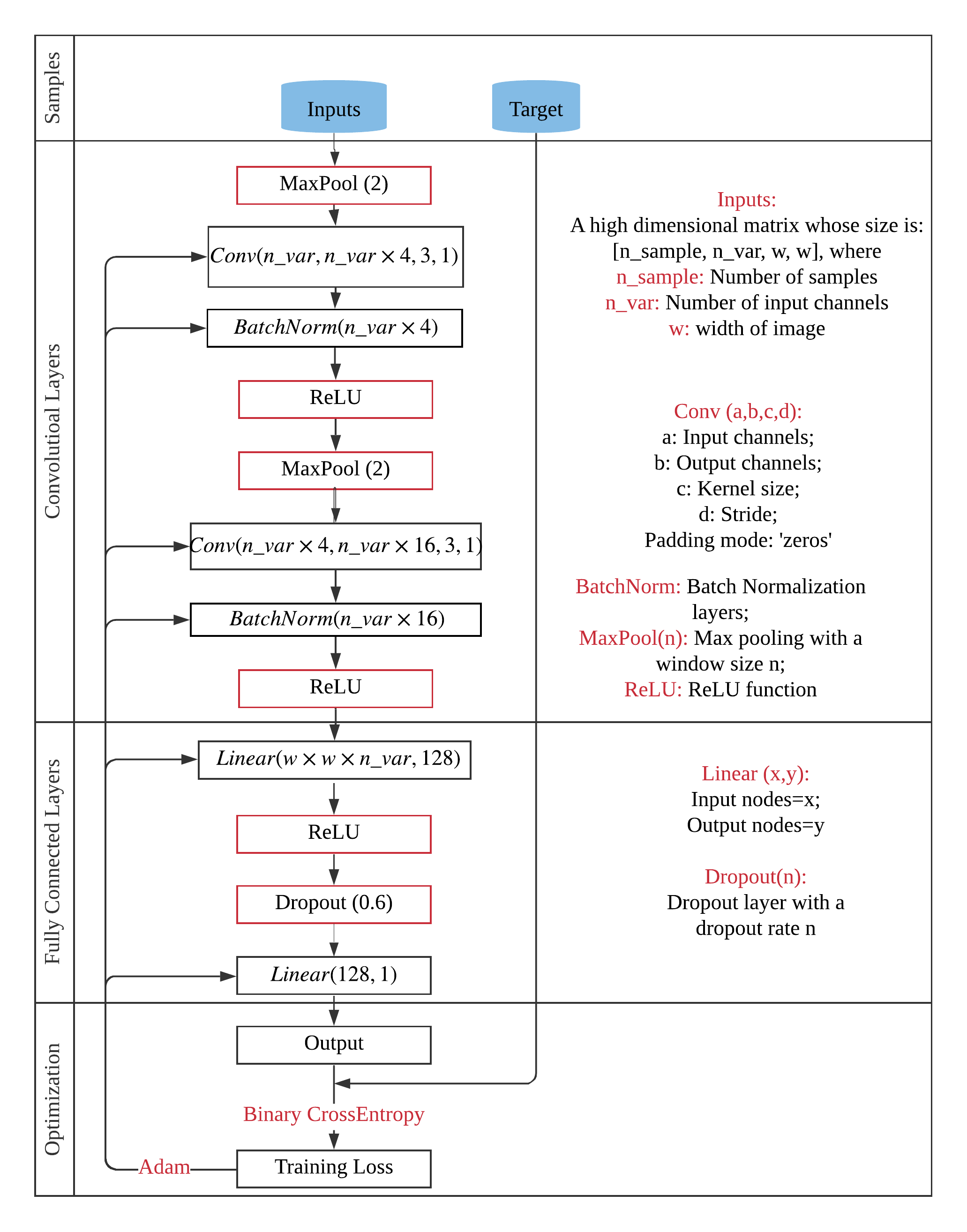}
  \caption{Architecture of CNN-X models.}
  \label{fig:cnn_architecture}
\end{figure}

As already mentioned, convolutional Neural Networks (CNNs) are widely used for image processing because their convolutional layers can extract essential information from images in a generic way. In this study, a standard CNN is used and implemented using PyTorch~\citep{NEURIPS2019_9015}. Each input channel is convolved with its own set of filters. The designed architecture of CNN-X model is shown in Fig. \ref{fig:cnn_architecture}.

\begin{figure}
  \centering
  \includegraphics[width=\textwidth]{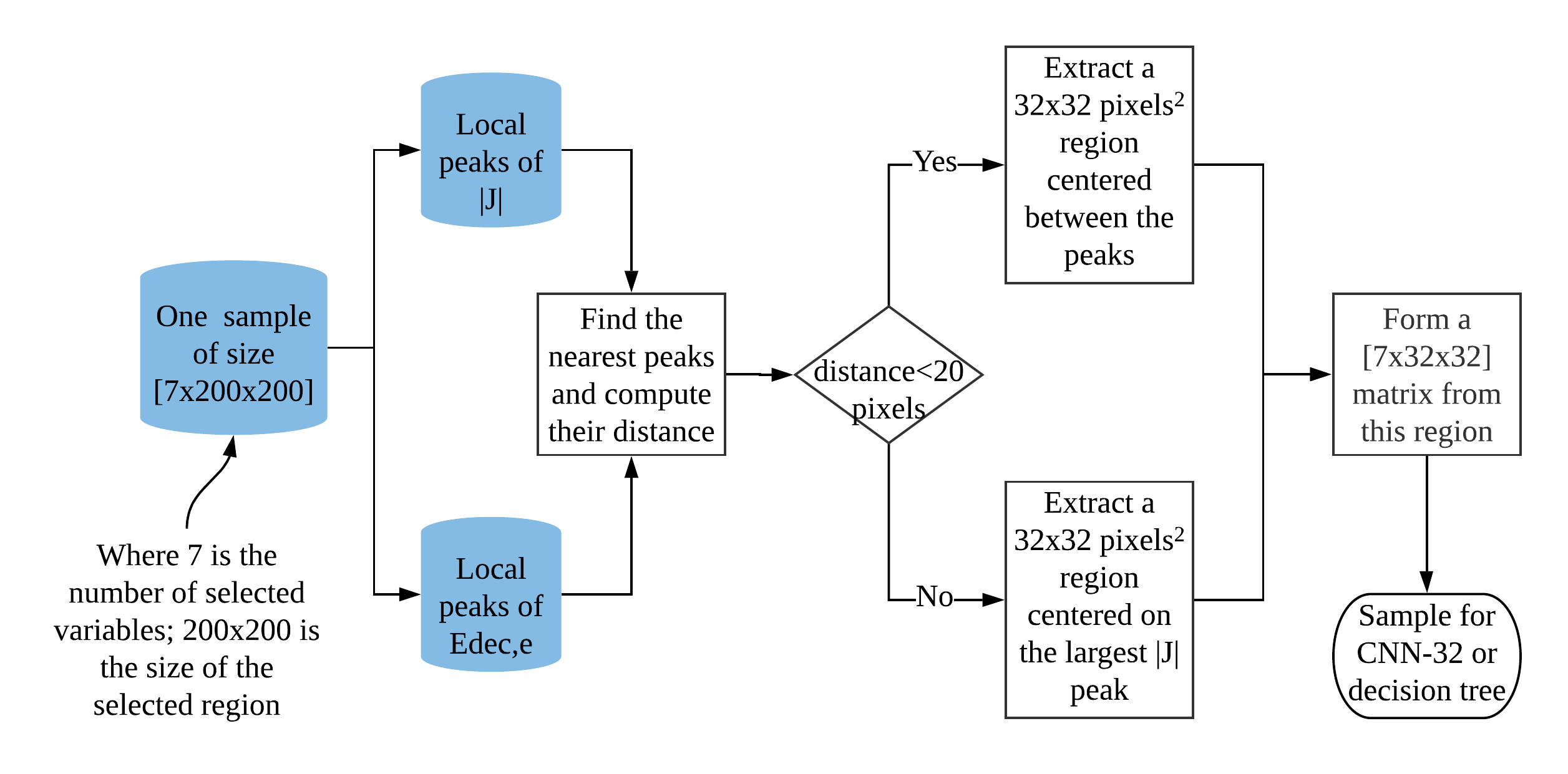}
  \caption{Flowchart of the image cropping approach, which extracts an $32^2$ pixel area from the $200^2$ pixel input.}
  \label{fig:feature-engineering}
\end{figure}

Because reconnection occurs when the dynamics of the magnetic field decouples from the electron fluid motion, there are two main signatures to consider: a peak in the current density $|{\bf J}|$ and a peak in the electron decoupling $E_{\mathrm{dec}, e}$ term (see Section \ref{sec:samples-labeling}). Based on these signatures, a heuristic method has been constructed to extract an area of $X^2$ pixels from the original $200^2$ input. In this study, 32 was found to be the optimal value for $X$, as discussed in Section \ref{sec:discussion-conclusion}. A flowchart describing this CNN-32 approach is shown in Fig. \ref{fig:feature-engineering}. It consists of the following steps:
\begin{enumerate}
  \item 
  For each image, find the closest pair of local maxima in $|{\bf J}|$ and $E_{\mathrm{dec}, e}$. This is done by first constructing a list of local maxima for each variable, consisting of all maxima that lie at least 10 pixels apart. Only local maxima whose amplitude is at least 70\% of the global maximum of the image are considered.
  \item If the closest pair is within a 20-pixel distance, extract a square region of $32 \times 32$ pixels centered on the middle of the two peaks. If not, return a $32^2$ region centered on the maximum of $|{\bf J}|$ in the image. All 7 physical variables are extracted.
\end{enumerate}
The resulting data set contains $7\times 32^2$ values (7 is the number of physical variables) per image, so it is about 40 times smaller than the original data set with $7 \times 200^2$ values per image. However, the $200^2$ images are still used for human labeling because they contain more information and are easier for human experts to label than the $32^2$ images.

\subsubsection{Decision tree classifier}
\label{sec:decision-tree}

This classifier takes as input only the $\min$, $\max$ and $\mathrm{mean}$ of each variable inside the $32^2$ area as illustrated in Fig. \ref{fig:feature-engineering}. Its input therefore consists of 21 variables (seven physical variables times three).
The decision tree classifier is implemented using the \textit{scikit-learn} library \citep{scikit-learn}, and its optimal depth is found with a grid search method.

\subsubsection{Optimal window size}
\label{subsubsec:win_size}
\begin{figure}
    \centering
    \includegraphics[width=0.6\textwidth]{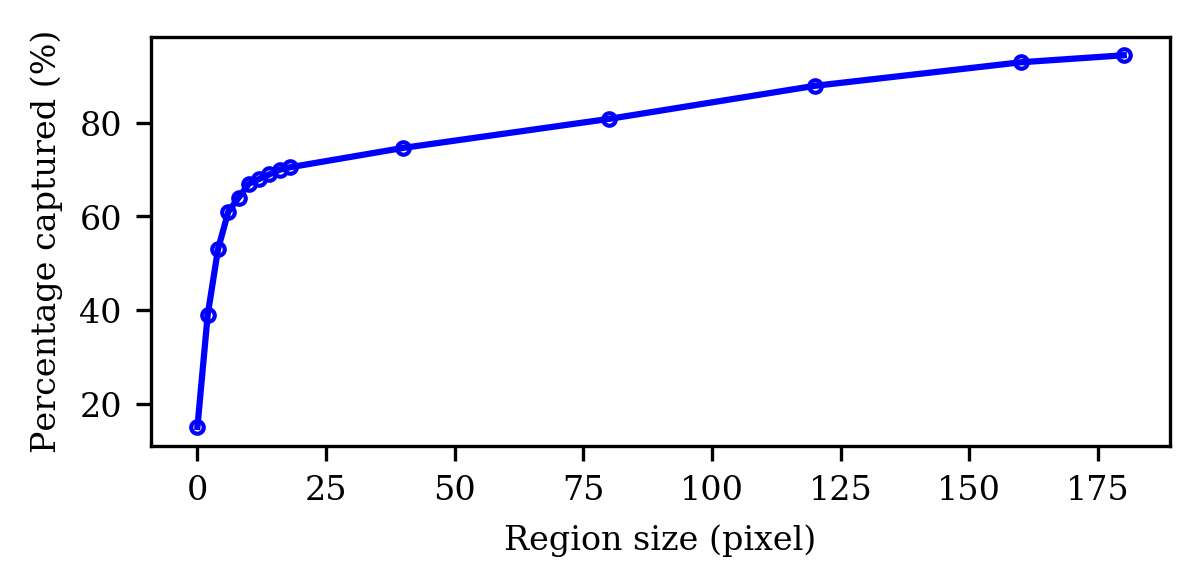}
    \caption{Percentage of reconnection sites captured versus window size for the image cropping approach described in section \ref{sec:feature-selection}.}
    \label{fig:win_acc}
\end{figure}

As introduced in Section \ref{sec:feature-selection}, a physics-based image cropping method is used in the CNN-X model and decision tree models to extract the most important $X^2$ pixel region out of the $200^2$ pixel input. When the ``correct'' region (i.e. the one containing the potential reconnection zone) is extracted, this approach increases the signal-to-noise ratio. A further benefit is that it reduces the computational cost of the CNNs, as the dimension of their input is reduced. However, the reconnection site marked by our human experts is not always captured inside the extracted region. Fig. \ref{fig:win_acc} shows the percentage of reconnection sites captured versus the window size of the extracted region. An ``elbow'' can be see around a size of about 20 pixels, at which size more than 70\% of reconnection sites can be captured. There are several reasons why not all reconnection sites are captured. For example, more than one reconnection site can be present in one sample, or it could be that no local maximum in $E_{\mathrm{dec}, e}$ can be found.

The accuracy of the models for different window sizes can be assessed by looking at the number of True Positives (TP), False Positives (FP), False Negatives (FN) and True Negatives (TN), where positives refer to reconnection events. To compare different models these numbers are converted to a single score. We consider two of such scores: Matthews' Correlation Coefficient (MCC) and the True Skill Statistic (TSS), which are defined as
\begin{align}
    \centering
    \label{eqn:mcc}
    \mathrm{MCC} &= \frac{\mathrm{TP} \times \mathrm{TN}  - \mathrm{FP} \times \mathrm{FN}}{\sqrt{(\mathrm{TP} + \mathrm{FP}) \times (\mathrm{FN} + \mathrm{TN}) \times (\mathrm{FP} + \mathrm{TN}) \times (\mathrm{TP} + \mathrm{FN})}}.\\
    \label{eqn:TSS}
    \mathrm{TSS} &= \mathrm{TPR}+\mathrm{TNR}-1, where \\
    \label{eqn:TPR}
    \mathrm{TPR} &= \frac{\mathrm{TP}}{\mathrm{FN} + \mathrm{TP}} \\
    \label{eqn:TNR}
    \mathrm{TNR} &= \frac{\mathrm{TN}}{\mathrm{FP} + \mathrm{TN}}
\end{align}

To be able to classify a potential reconnection site, it can be beneficial to have information available in a neighborhood around the site. So there are two competing factors: a smaller window size can increase the signal-to-noise ratio whereas more reconnection sites and more of their surroundings can be captured with a larger window size. To determine the optimal window size, we compare the performance of the CNN-16, CNN-32, CNN-64, CNN-128 and CNN-200 models on the validation set described in table \ref{tab:data sets}. The TSS, MCC and confusion matrix of these models are shown in Table \ref{tab:performance-in}. The listed numbers are average values obtained by training the model 10 times with the same configuration but different initial random values for the CNN weights. Table \ref{tab:performance-in} shows that the optimal window size is 32 for this application.
The ``image cropping'' method described in section \ref{sec:ml-methods} can therefore improve both the accuracy and efficiency of the CNN-X model. With a larger window size, the performance degrades because there is more noise in the input, such as other potential reconnection sites or complex structures due to turbulence. This noise makes it more difficult to train a CNN that performs well on the test set. We remark that with a larger data set a larger window size could work better because noise would be less of a concern.

With a window size of 32, the fraction of cases that is misclassified is about 26\%. For reference, we expect that about 14\% of cases in the data set would be marked ambiguous if they all had been labeled by three human experts, see section~\ref{sec:samples-labeling}.

\begin{table}[!htbp]
\centering
\caption{Accuracy of the CNN-X models with different window sizes X, evaluated on the out-of-sample Sim 1 data. The results shown are averages over 10 trained models with different initial random coefficients.}
\begin{tabular}{c | c c | c c c c}
  \hline
  Window size & TSS & MCC & TP & FP & TN & FN\\
  \hline
  16 & 0.29 & 0.32 & 85 & \bf{12} & \bf{176} & 158 \\
  32 & \bf{0.56} & \bf{0.55} & \bf{170} & 28 & 161 & \bf{70} \\
  64 & 0.42 & 0.41 & 138 & 29 & 159 & 103 \\
  128 & 0.43 & 0.44 & 133 & 23 & 166 & 108 \\
  200 & 0.39 & 0.44 & 154 & 48 & 141 & 87 \\
\end{tabular}
\label{tab:performance-in}
\end{table}

\section{Results}
\label{sec:results}

In this section, we evaluate the accuracy of the two machine learning approaches introduced in section \ref{sec:ml-methods}, namely the decision tree model and the CNN-32 model, which was found to be optimal in section \ref{subsubsec:win_size}. The training set introduced in Table \ref{tab:data sets} is used to train these models. 
In section \ref{sec:accuracy}, the test set, which is Sim 2, is used to assess the accuracy of both models. In section \ref{subsubsec:feat_sel}, the importance of the physical variables is investigated by comparing the accuracy of the CNN-32 model on the test set with different inputs.

\subsection{Model accuracy}
\label{sec:accuracy}

The scores of the CNN-32 model (which is the optimal CNN-X model) and the decision tree model are shown in Table \ref{tab:performance-out}. Both models are evaluated on the test set described in Table \ref{tab:data sets}. As before, the score of the CNN-32 model is averaged over 10 model instances. The CNN-32 model significantly outperforms the simple decision tree classifier. It has a true positive rate (TPR or sensitivity, details in Eqn. \ref{eqn:TPR}) of $0.82$ and a true negative rate (TNR or specificity, details in Eqn. \ref{eqn:TPR}) of $0.66$. For the decision tree, these rates are $0.67$ and $0.63$, respectively.

The performance of the CNN-32 model on the test set is almost as good as on the out-of-sample validation data (Table \ref{tab:performance-in}). This indicates that the model can be applied to independent simulations performed with the same numerical code, provided that the grid resolution is not too different. To apply the model to a simulation performed at a very different resolution, some scaling of the input would probably be required, so that a $32^2$ pixel window would capture a region of similar physical size.
In future work, it would also be interesting to investigate the performance of the CNN-32 model on 2D reconnection simulations performed with different numerical codes.

\begin{table}[!htbp]
\centering
\caption{Accuracy of the machine learning models evaluated on test data set, see Table \ref{tab:data sets}. }
\begin{tabular}{c | c c | c c c c}
  \hline
  Model & TSS & MCC & TP & FP & TN & FN\\
  \hline
  CNN-32 & 0.50 & 0.51 & 48 & 10 & 43 & 22 \\
  Decision tree & 0.28 & 0.30 & 55 & 27 & 26 & 15 \\
\end{tabular}
\label{tab:performance-out}
\end{table}

\subsection{Importance of physical variables to classifier accuracy}
\label{subsubsec:feat_sel}

\begin{table}
\centering
\caption{Performance of CNN-32 models when some of the variables are randomly shuffled within the test set, meaning they are no longer informative of reconnection. The numbers in each row indicate how much the MCC score improves when the corresponding variables are unshuffled. The most important variable per row is marked in red. These variables are included unshuffled in the rows below. Improvements in the MCC score are shown in the rightmost column. The reported numbers are averages over 10 CNN-32 model instances.}
\begin{tabular}{c | c c c c c c c | c}
  \hline
    & & & & MCC &  & & & \\
  No. & $|{\bf J}|$ & $B_\mathrm{plane}$ & $V_{e,z}$ & $E_{\mathrm{dec}, e}$ & $\delta B_z$ & $\Psi$ & $V_{e,\mathrm{plane}}$ & Improvement\\
  \hline
    1 & \color{red}{0.33} & 0.051   & 0.16 & 0.024   & 0.001 & -0.044 & 0.015    &  0.33\\
    \hline
    2 &  \checkmark   & \color{red}{0.43} & 0.39 & 0.32 & 0.34 & 0.33 & 0.35 & 0.098\\
    \hline
    3 &  \checkmark   &  \checkmark   & \color{red}{0.49} & 0.40 & 0.42 & 0.39 & 0.42 & 0.062\\
    \hline
    4 &  \checkmark   &  \checkmark   &  \checkmark   & \color{red}{0.50} & 0.48 & 0.40 & 0.43 & 0.007\\
    \hline
    5 &  \checkmark   &  \checkmark   &  \checkmark   &  \checkmark   & \color{red}{0.50} & 0.49 & 0.48 & 0.004\\
    \hline
    6 &  \checkmark   &  \checkmark   &  \checkmark   &  \checkmark   &  \checkmark   & \color{red}{0.50} & 0.49 & 0.003\\
    \hline
    7 &  \checkmark   &  \checkmark   &  \checkmark   &  \checkmark   &  \checkmark   &  \checkmark   & \color{red}{0.50} & 0.0\\
\hline
\end{tabular}
\label{tab:Feat_analysis_model}
\end{table}

One of the objectives of this study is to investigate which physical variables are important for the classification of magnetic reconnection. We do this in two ways. The first approach is to shuffle the variables in the test set, and then determine a sequential order for their importance as follows:
\begin{enumerate}
    \item Train the CNN-32 model 10 times to have 10 different CNN-32 models.
    \item Shuffle each physical variable in the test set, by randomly permuting the corresponding $32^2$ pixel data over the samples.
    \item Put the original values back in the test set for one variable at time in order to check the influence of each variable on the classification.
    \item Find the variable which has the largest influence (largest mean MCC score in 10 models) and `freeze' it, meaning that it will not be shuffled anymore. Then go back to step 3 to test other variables, until no variables are left.
\end{enumerate}

The results are shown in Table \ref{tab:Feat_analysis_model}. They reveal that, statistically, $|{\bf J}|$ contributes most in this classification model. $B_\mathrm{plane}$ and $V_{e,z}$ are slightly less significant. The other variables $E_{\mathrm{dec}, e}$, $\delta B_z$, $\Psi$ and $V_{e,\mathrm{plane}}$ are not very significant, as the MCC score improves by less than 1\% when one of them is added. 

Physically, these results seem reasonable because 1) $|{\bf J}|$, $B_\mathrm{plane}$ and $V_{e,z}$ are all directly related to the reconnection process, and they are highly correlated; 2) $E_{\mathrm{dec}, e}$ has already been used for the image cropping method (as introduced in Section \ref{sec:feature-selection}) and 3) $V_{e,\mathrm{plane}}$ variations are a consequence of reconnection, but they can also be caused by turbulence. Overall, Table \ref{tab:Feat_analysis_model} indicates that the first three variables, $|{\bf J}|$, $V_{e,z}$, $B_\mathrm{plane}$, might be enough to develop a reconnection classification model as good as the one based on all seven variables.

The above results say something about the importance of variables in a model that was trained on all variables. A slightly different question is which variables are the most important when a model with fewer inputs is used. To investigate this we have trained CNN-32 models using one, two or three physical variables as input. All possible combinations of variables were tested. Table \ref{tab:combinations} shows the MCC scores of the top five combinations. Again, $|{\bf J}|$ is the most important variable followed by $B_\mathrm{plane}$ and $V_{e,z}$, in agreement with the results from Table \ref{tab:Feat_analysis_model}.

\begin{table}
\centering
\caption{MCC scores of CNN-32 models that only take the listed variables as input, evaluated on the test set. The best five combinations of variables are shown for one, two or three physical input variables. The results are averages over 10 model instances.}
\begin{tabular}{lc|lc|lc}
  \hline
  1 variable & MCC & 2 variables & MCC & 3 variables & MCC\\
  \hline
  $|{\bf J}|$ & 0.44 & $|{\bf J}|$, $B_\mathrm{plane}$ & 0.51 & $|{\bf J}|$, $V_{e,z}$, $B_\mathrm{plane}$ & 0.56 \\
  $V_{e,z}$ & 0.39 & $|{\bf J}|$, $V_{e,z}$ & 0.49 & $|{\bf J}|$, $B_\mathrm{plane}$, $E_{\mathrm{dec}, e}$ & 0.55 \\
  $V_{e, plane}$ & 0.13 & $|{\bf J}|$, $E_{\mathrm{dec}, e}$ & 0.44 & $|{\bf J}|$, $V_{e,z}$, $\Psi$ & 0.50 \\
  $B_\mathrm{plane}$ & 0.11 & $|{\bf J}|$, $\Psi$ & 0.42 & $|{\bf J}|$, $\Psi$, $E_{\mathrm{dec}, e}$ & 0.50 \\
  $\delta B_z$ & 0.09 & $|{\bf J}|$, $V_{e, \mathrm{plane}}$ & 0.40 & $|{\bf J}|$, $V_{e,z}$, $E_{\mathrm{dec}, e}$ & 0.48
\end{tabular}
\label{tab:combinations}
\end{table}

\section{Discussion}
\label{sec:discussion-conclusion}

\subsection{Analysis of wrong predictions}
\label{subsec:False_ana}

\begin{figure}
\includegraphics[width=\textwidth]{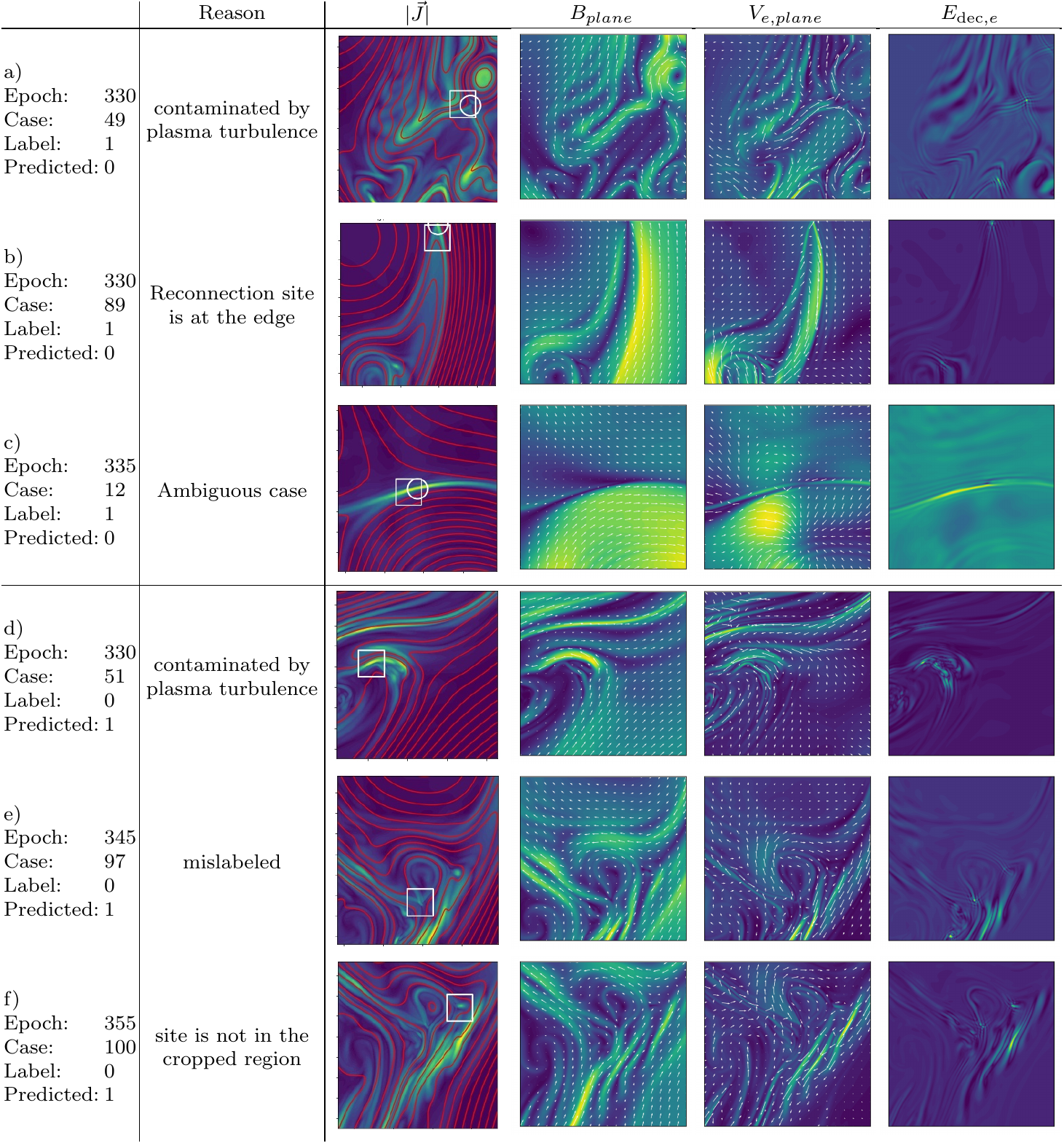}
\caption{Examples of false negatives (rows a-c) and false positives (rows d-f). Reconnection sites labeled by human experts are indicated by white circles. The region extracted by the image cropping method is shown by white squares. Possible reasons for the false predictions are indicated. The images show the same variables as in Fig. \ref{fig:Mask}, in a $200^2$ pixel window. }
\label{fig:FN&FP}
\end{figure}

In this section, we investigate why the CNN-32 model in some cases makes a wrong prediction. Three examples of false negatives (FN) and three examples of false positives (FP) are shown in Fig. \ref{fig:FN&FP}. Here, a FP refers to a case labeled 0 (no reconnection) for which the model predicts a label 1 (reconnection), and a FN is a case labeled 1 for which the predicted label is 0. Each panel is a ``candidate'' magnetic reconnection event similar to Fig. \ref{fig:Mask}, where panels a-c display FN cases and panels d-f display FP cases. 

Each case illustrates a different type of mis-classification. In Fig. \ref{fig:FN&FP}a, a so-called plasmoid is present with a relatively complex structure around it. However, the reason the labeled reconnection point was not recognized is probably that it does not coincide with a significant peak in the current density. The reconnection site in Fig. \ref{fig:FN&FP}b is at the edge of the image. Such sites are probably harder to classify since only ``half'' of the reconnection zone is visible. Case c) is ambiguous, even for human experts. For the false positives, Case d) is distorted by plasma turbulence and, as a consequence, has a complex morphology. Finally, cases e) \& f) are actually correctly predicted but have a wrong label. These cases correspond to the same physical region at different times. It is clear that reconnection proceeds at the lower edge of the magnetic flux rope in the center of case e), where an X-point is clearly visible. Indeed the difference between the $\Psi$ value at the X-point and its value at the O-point (the center of the flux rope) increases in time passing from e) to f). The fact that the main current peak is away from the X-point misled humans during classification, but the machine learning model correctly catches reconnection going on. However, in case f), the cropping method in this particular case selects a wrong region. This is due to the presence of a high peak in the current density and in $E_\mathrm{dec}$ at some distance from the true reconnecting site.

These cases show that the CNN-32 model is in some cases able to find reconnection sites that are initially missed by human experts.

\section{Summary \& Outlook}
\label{sec:summary-outlook}

The first extensive labeled data set for magnetic reconnection in 2D HVM simulations has been constructed with currently over 2000 samples labeled by human experts. We have developed a classifier able to automatically identify magnetic reconnection in such simulations using convolutional neural networks (CNNs). An important part of this classifier is a physics-based image-cropping method that zooms in on potential reconnection sites. The overall model is called `CNN-X', where X indicates the size of the cropped window. Our results show that:

\begin{enumerate}
    \item The image cropping method can improve the accuracy of the CNN models by increasing the signal-to-noise ratio. The optimal window size around potential reconnection sites was found to be $32^2$ pixels. The corresponding CNN-32 model had a true positive rate (sensitivity) of 89\% and a true negative rate (specificity) of 70\% when evaluated on an out-of-sample validation set.
    \item The CNN-32 model was also evaluated on a fully independent test set that was constructed from a simulation with lower resolution. The model then had a true positive rate of 82\% and a true negative rate of 66\%. This indicates the developed CNN-32 model is generic and can be applied to other simulations. Furthermore, in some cases, the CNN-32 model was able to find reconnection sites that were initially missed by a human expert.
    \item We have investigated the importance of different physical variables for the predection of reconnection. Three variables were found to be the most important reconnection markers: the current density $|{\bf J}|$, the out-of-plane electron velocity $V_{e,z}$ and the in-plane magnetic field $B_\mathrm{plane}$.
\end{enumerate}

This study is a first step in adopting machine learning for the automatic identification of magnetic reconnection. We think that with more labeled data from different types of simulations the model's accuracy would improve. This would then open up the possibility of using machine learning to detect reconnection in other types of data, such as artificial and real satellite measurements. In particular, the long-term goal of this study is to use the the classifications obtained using the model developed here to analyze time series data created using a virtual satellite technique. These time series could then be collected in a large labeled database, which would help to detect reconnection in time series recorded by real satellites.

\acknowledgments

This project has received funding from the European Union's Horizon 2020 research and innovation programme under grant agreement No 776262 (AIDA, www.aida-space.eu)
Numerical simulations have been performed on Marconi at CINECA (Italy) under the ISCRA initiative. FC thank Dr.~M.~Guarrasi (CINECA, Italy) for his essential contribution on code implementation on Marconi.

The simulation dataset (UNIPI\_TURB\_2D, UNIPI\_TURB\_2D\_2048) is available at Cineca on the AIDA-DB. In order to access the meta-information and the link to the raw data, look at the tutorial at http://aida-space.eu/AIDAdb-iRODS.


This publication uses data generated via the \url{Zooniverse.org} platform, development of which is funded by generous support, including a Global Impact Award from Google, and by a grant from the Alfred P. Sloan Foundation.

For their help in labeling the data set we want to thank: Sid Fadanelli, Giuseppe Arr\`o, Giulia Cozzani, Silvio Sergio Cerri, Francesco Pucci, Francesco Pegoraro, Alessandro Retin\`o, Oreste Pezzi, J\"org B\"uchner, Amir Chatraee, and Neeraj Jain.

\textbf{The current version of the labeled data set and the code for the machine learning models are available on Zenodo (\url{https://doi.org/10.5281/zenodo.3907309} and \url{https://doi.org/10.5281/zenodo.3935887} respectively). Please check the AIDA website at \url{http://aida-space.eu/reconnection} for further updates.}

\bibliography{reconnection}
\bibliographystyle{aasjournal}



\end{document}